\magstep1
\parindent =0pt
\def\SU{\mathcode`,="8000 \mathcode`!="8000\relax}      
{\catcode`\,=\active \catcode`\!=\active
\gdef,{\mkern-2mu{}_} \gdef!{\mkern-2mu{}^}}
\def\Ko{\oalign{$K$\cr  
        \noalign{\vskip0.3ex}   
        \hidewidth$\scriptstyle0$\hidewidth}{}}
\def\Ro{\oalign{$R$\cr
\noalign{\vskip0.3ex}
\hidewidth$\scriptstyle0$\hidewidth}{}}
\def\Po{\oalign{$P$\cr
\noalign{\vskip0.3ex}
\hidewidth$\scriptstyle0$\hidewidth}{}}
\baselineskip18pt

\def\c {$\SU C_{abcd}$ }
\def\w {$\SU W_{abcd}$ }
\def\h {$\SU L_{abc}$ }

\def\Po{$\SU \Po^a,{bcd}$ }

\def\Ko{$\SU \Ko^a,{bcd}$ }

\def\Ro{$\SU \Ro^a,{bcd}$ }

\def\square{\,\hbox{\vrule\vbox{\hrule\phantom{N}\hrule}\vrule}\,}

\def \NR {Novello and Velloso } \def \NV {Novello and Velloso }

\def \BC {Bampi and Caviglia }

\def \II {Illge }

\def \BS {Bell and Szekeres }

\def \DK {Dolan and Kim }

\def \BS {Bell and Szekeres }

\def \frac#1#2{{#1 \over #2}}

\parindent=0pt

\vskip.5in
\centerline{\bf    THE LANCZOS POTENTIAL FOR THE WEYL CURVATURE TENSOR:
 }
\centerline{\bf     EXISTENCE, WAVE EQUATION AND ALGORITHMS.
 }

\

\centerline{S. Brian Edgar and A. H\"oglund}

\centerline{ \it Department of Mathematics, 
 Link\"oping University,}

\centerline{\it Link\"oping,} 

\centerline{\it S581 83 Sweden. }

\

\beginsection ABSTRACT

In the last few years renewed interest in the 3-tensor potential $L_{abc} $
proposed by Lanczos for the Weyl curvature tensor has not only clarified and
corrected Lanczos's original work, but generalised the concept in a number of
ways. In this paper we carefully summarise and extend some aspects of these
results, and clarify some misunderstandings in the literature.  We also clarify
some comments in a recent paper by Dolan and Kim; in addition, we correct some
internal inconsistencies in their paper and extend their results.

The following new results are also presented. The (computer checked) complicated
second order partial differential equation for the 3-potential, in arbitrary
gauge, for Weyl candidates satisfying Bianchi-type equations is given --- in
those $n $-dimensional spaces (with arbitrary signature) for which the potential
exists; this is easily specialised to Lanczos potentials for the Weyl curvature
tensor. It is found that it is {\it only} in 4-dimensional spaces (with
arbitrary signature and gauge), that the non-linear terms disappear and that the
awkward second order derivative terms cancel; for 4-dimensional spacetimes (with
Lorentz signature), this remarkably simple form was originally found by Illge,
using spinor methods.  It is also shown that, for most 4-dimensional vacuum
spacetimes, any 3-potential in the Lanczos gauges which satisfies a simple
homogeneous wave equation must be a Lanczos potential for the Weyl curvature
tensor of the background vacuum spacetime. This result is used to prove that the
form of a {\it possible} Lanczos potential proposed by Dolan and Kim for a class
of vacuum spacetimes is in fact a genuine Lanczos potential for these
spacetimes.

\vfill\eject

\beginsection 1. Introduction

Although the existence of a 3-tensor $L_{abc} $ as a potential for the Weyl
conformal tensor $C_{abcd} $\footnote{${}^{\dag}$} {In general, for tensors and
spinors, we follow the notation and conventions of Penrose and Rindler (1984);
for the Lanczos tensor/spinor we follow the notation of Dolan and Kim (1994a,b)
(except for the definitions of the tetrad/dyad components of the Lanczos tensor,
as we shall explain in Section 3).  We emphasise the importance of care with
notations and conventions; this is the cause of some of the previous and present
misunderstandings. In particular we point out that in each one of the papers to
which we refer in some detail --- Lanczos (1962), Bampi and Caviglia (1983),
Illge (1988), Novello and Velloso (1987), Zund (1975), Ares de Parga G., et al
(1989) --- at least one different convention is used compared to those used in
this paper.}  in 4-dimensional spacetimes was first suggested by Lanczos (1962),
for some time there was little significant development --- probably due to some
mistakes and misunderstandings, both in the original paper, and in some
subsequent papers. However, more recently a number of interesting results have
been obtained. There are a number of subtleties in these results --- involving
dimension, signature, gauge, and indeed the class of tensors for which
potentials can be found --- and again a number of misunderstandings and mistakes
have crept into the literature; so we shall first present a careful summary of
these results.

The topic was placed on a firm foundation and in a wider context by Bampi and
Caviglia (1983); they detected a logical flaw in Lanczos's original argument,
and not only gave a valid and rigorous proof of existence, but extended
Lanczos's original proposal to a larger class of 4-tensors, to a larger class of
3-tensors, and to a larger class of spaces.  \BC have shown --- in any
4-dimensional analytic manifold with metric $g_{ab} $ (irrespective of
signature) --- that any analytic 4-tensor \w with the properties
     $$\eqalign{W_{abcd} = W_{[ab]cd} & =W_{ab[cd]}\cr 
       W^a{}_{bad}= 0 & = W_{a[bcd]}}\eqno(1.1)$$
always admits locally a regular analytic 3-tensor potential \h with the
properties, 
  $$\eqalign{ L_{abc}= & L_{[ab]c}\cr L_{[abc]}& =0} \eqno(1.2) $$
according to
$$\eqalign{ W_{abcd} = & \  2 L_{ab[c;d]}
+2L_{cd[a;b]}  - g_{a[c}(L_{|b|}{}^e{}_{d];e}-L_{|b|}{}^e{}_{e;d]}+
L_{d]}{}^e{}_{b;e}-L_{d]}{}^e{}_{e;b}) \cr & +g_{b[c}(
L_{|a|}{}^e{}_{d];e}-L_{|a|}{}^e{}_{e;d]}+
L_{d]}{}^e{}_{a;e}-L_{d]}{}^e{}_{e;a})  +{4\over 3}
g_{a[c}g_{d]b}L^{ef}{}_{e;f}}  \eqno(1.3) $$
(The original presentation by Lanczos (1962) was not in the form (1.3) but
rather in an equivalent form using the Hodge dual operator.)

We shall refer to a tensor \w
with the properties (1.1) as a {\it Weyl conformal  curvature tensor candidate}
(or {\it Weyl tensor candidate}); and obviously any 3-tensor $L_{abc} $,
satisfying (1.2), acts as a potential for some Weyl  tensor candidate. 

Lanczos had imposed the
additional conditions,
$$L_{ab}{}^b=0 \qquad\qquad \hbox{(a)}\qquad \qquad 
L_{ab}{}^c{}_{;c}=0 \qquad\qquad \hbox{(b)}\eqno(1.4)$$
and although he had shown directly the admissibility of the first, he only
provided a heuristic argument for the admissibility of the second.  On the other
hand, \BC showed that these restrictions are not essential, and gave a rigorous
proof for the admissibility of any values on the right hand sides of the two
equations in (1.4).  Since these choices have no effect on the Weyl candidate
defined by (1.3) they are gauge choices, and (1.4a,b) are referred to
respectively as the {\it Lanczos algebraic gauge} and the {\it Lanczos
differential gauge}.\footnote{${}^{\dag}$}{The status of these conditions is not
always made explicit.  In Ares de Parga et al. (1989) and L\'opez-Bonilla et
al. (1993) the Lanczos algebraic gauge (1.4a) is used as part of the definition
of the potential with no mention that it is a gauge condition, but the
differential gauge condition is not mentioned; while the set of equations (1.3)
is given in an abreviated form, with the Lanczos algebraic gauge condition
(1.4a) substituted.  In Dolan and Kim (1994a,b) the condition (1.4a) is also
used as part of the definition of the potential with no mention that it is a
gauge condition, but the Lanczos differential gauge condition (1.4b) is used,
and explicitly described as such; while the set of equations (1.3) is quoted in
the most general form without any gauge simplifications. In spinor
presentations, the Lanczos algebraic gauge condition is built into the symmetry
of the usual Lanczos spinor $L_{ABC\dot A} =L_{(ABC)\dot A} $, but it would be
easy to define a different spinor with an arbitrary algebraic gauge; the
differential gauge is not built in, being given explicitly by (1.6), but its
imposition means that (1.5) can be written as $\eqalign{W_{ABCD} =
2\nabla_{A}{}^{\dot A} L_{BCD \dot A} }$}.

   Illge (1988) has supplied a much simpler existence proof than that in Bampi
and Caviglia (1983); however, since this proof is obtained using the spinor
formalism of Penrose and Rindler (1984), it is valid only in spacetimes
(i.e. four dimensional spaces with {\it Lorentz} signature).  He has shown that
any symmetric 4-spinor $W_{ABCD} $ ({\it Weyl spinor candidate}) admits locally
a solution to
   $$\eqalign{W_{ABCD} = 2\nabla_{(A}{}^{\dot A} L_{BCD) \dot A} 
}\eqno(1.5)$$
where $ L_{BCD\dot A}$ is
symmetric in all undotted indices (which means that the Lanczos algebraic gauge
is built in). If in addition $ L_{BCD\dot A}$  satisfies the
Lanczos   differential gauge  
 $$ \nabla^{A \dot A} L_{ACD \dot A}=0 
\eqno(1.6)$$
 then the symmetry brackets in (1.5) may be omitted.

  It is
also shown  in  Illge (1988)\footnote{${}^{\dag}$}{ We  emphasise that the
results
 in  Illge (1988) are much more general than the extract summarised here: the
main theorem  is actually given in terms of a Cauchy problem, and both the
theorem and the wave equation obtained there are for the  case of a symmetric
$n$-spinor $W_{ABCD . . . N} $ in general non-vacuum 4-dimensional spacetimes,
with spinor potential in arbitrary differential gauge.}  that if $W_{ABCD} $
satisfies a homogeneous Bianchi-type equation of the form
$$\eqalign{
\nabla^{A \dot A}W_{ABCD} =0}\eqno(1.7)$$
then $L_{BCD\dot A} $ satisfies a homogeneous wave equation, which in vacuum
(i.e. Einstein tensor $G_{ab}=0$) and in both Lanczos gauges has the  remarkably
simple form  $$\eqalign{
\square L_{ACD \dot A}=0}\eqno(1.8a)$$
or equivalently in tensor notation in both Lanczos gauges, 
$$\eqalign{
\square L_{abc}=0}\eqno(1.8b)$$
where $\square (=\nabla_a\nabla^a) $ is the differential wave operator in four
dimensional spacetime. 

However, when the analogous analysis is carried through in
tensor notation, i.e.  for a  potential tensor $L_{abc} $  satisfying Lanczos
gauges  (1.4), and for a Weyl candidate $W_{abcd} $ satisfying a homogeneous 
Bianchi-type equation
 $W_{abc}{}^d{}_{;d}=0 $ in vacuum (i.e. Einstein tensor $G_{ab}=0$), we obtain
$$\eqalign{ \nabla^2 L_{abc}= 2L^{def}g_{[a|c|}C_{b]def}
-2L_{[a}{}^{de}C_{b]edc}-{1\over2} L^{de}{}_c C_{deab}}\eqno(1.9)$$
where $\nabla^2 (=\nabla_a\nabla^a)$ and $C_{abcd} $ is the Weyl tensor of the
background four dimensional space of arbitrary signature.  This appears, at
first sight, to be much more complicated than the homogeneous wave equation
(1.8b), for the special case of Lorentz signature. However, this apparent
disagreement has been resolved in Edgar (1994a) where it is shown that the
expression on the right hand side of (1.9), is identically zero in four (and
only four) dimensions i.e.
 $$\eqalign{
 2L^{def}g_{[a|c|}C_{b]def}
-2L_{[a}{}^{de}C_{b]edc}-{1\over2} L^{de}{}_c C_{deab}\equiv0}\eqno(1.10)$$
where $L_{abc}
$ is in the algebraic Lanczos gauge.  It is emphasised that this
identity is a consequence of purely algebraic properties; no 
differential properties are assumed between $L_{abc} $ and $C_{abcd} $.
\footnote{${}^{\dag}$}{In fact 
this is a special case of a more general 4-dimensional identity, 
$$\eqalign{
 2L^{def}g_{[a|c|}W_{b]def}
-2L_{[a}{}^{de}W_{b]edc}-{1\over2} L^{de}{}_c W_{deab} - {1\over 2}
W_{abcd}L^{de}{}_e\equiv 0}$$ 
 where $L_{abc} $ is in arbitrary algebraic  gauge,  and $W_{abcd}
$ is any Weyl candidate, (Edgar and H\"oglund (1995)).} So therefore, in four
dimensional spaces of any signature, (1.8b) generalises to 
$$\nabla^2L_{abc}=0
\eqno(1.8c)$$

\medskip 
Clearly, since the Weyl conformal  tensor
\c and its spinor counterpart $\Psi_{ABCD} $, are special cases of the
Weyl candidates discussed above,  the above results for existence of a
potential apply directly to them.   We shall
reserve the term {\it Lanczos potential } for   the potential of a Weyl
conformal curvature tensor \c (or spinor $\Psi_{ABCD} $) which can therefore
also be written, in arbitrary gauges, in four dimensions with arbitrary
signature as    $$\eqalign{ C_{abcd} = & \  2 L_{ab[c;d]}
+2L_{cd[a;b]}  -
g_{a[c}(L_{|b|}{}^e{}_{d];e}-L_{|b|}{}^e{}_{e;d]}+
L_{d]}{}^e{}_{b;e}-L_{d]}{}^e{}_{e;b})
\cr & +g_{b[c}(
L_{|a|}{}^e{}_{d];e}-L_{|a|}{}^e{}_{e;d]}+
L_{d]}{}^e{}_{a;e}-L_{d]}{}^e{}_{e;a})  +{4\over
3} g_{a[c}g_{d]b}L^{ef}{}_{e;f}} 
\eqno(1.11a) $$
Since in vacuum
the Weyl tensor satisfies the Bianchi identity in the form, 
$$\eqalign{\nabla^d C_{abcd}=0 }\eqno(1.12a)$$
the
second order equation (1.8c) is also valid, in vacuum,  for the Lanczos
potential of the  Weyl  tensor $C_{abcd} $.

The Weyl spinor  is given by
 $$\eqalign{\Psi_{ABCD} =
2\nabla_{(A}{}^{\dot A} L_{BCD) \dot A}  }\eqno(1.11b)$$
in  Lanczos algebraic gauge, and since in vacuum
the Weyl spinor  satisfies the Bianchi identity in the form, 
$$\eqalign{\nabla^{A\dot A}\Psi_{ABCD} = 0 
}\eqno(1.12b)$$
the
wave equation (1.8a)\footnote{$^{\dag}$}{Although  Lanczos (1962) had
derived a wave equation for his potential of the Weyl tensor, in four
dimensional spacetimes, he had  misplaced some of his indices and so his version
differs from (1.9); this version has been repeated uncorrected by Atkins and
Davis (1980), and has been only  partly corrected by Roberts (1989). But it was
the work of Illge (1988) {\it in spinors} that revealed the very simple form
(1.8) of this wave equation, and gave the hint to the existence of the
4-dimensional identity (1.10).} is also valid, in vacuum,  for the Lanczos
potential of the  Weyl spinor
$\Psi_{ABCD} $.
\medskip
In a recent paper Dolan and Kim (1994a) also derive the tensor identity (1.10)
in four dimensional spacetimes by {\it spinor methods}, and this leads them to
make the claims that this identity `is restricted to Lorentz signature' and also
`that spinor methods seem to be essential to prove all the 4-dimensional
identities'.  Although \DK do refer to the proof for this identity given in
Edgar (1994a), they seem to have overlooked the fact that, in that paper, a {\it
tensor} proof is given, which is explicitly stated and shown to be {\it
independent of metric signature}; in fact the identity was not verified by the
computer algebra system STENSOR as Dolan and Kim state, but rather by a simple
tensor manipulation with Hodge duals for four dimensional spaces.  In addition,
\DK have given another identity --- (D1) in Dolan and Kim (1994a) --- which they
also claim to be `strictly four-dimensional and restricted to Lorentz
signature'; but this identity is easily seen to be a direct consequence of the
identity (1.10), and so once again signature is irrelevant. \footnote{$^{\ddag}
$}{These are not the only identities with dimensionally dependent properties
which arise in this manner (see Andersson and Edgar (1995)) nor the only time
that spinor methods have led to the mistaken conclusion that the signature has
some relevance, (see Dianyan (1986)); there has also recently been a related
discussion in the context of algebraic invariants of the Riemann tensor, by Jack
and Parker (1987), Fulling et al. (1992) and Harvey (1995).  Colleagues have
drawn our attention to an earlier paper by Lovelock (1970) who shows explicitly
that such types of identities are a trivial, but subtle, consequence {\it of
dimension alone}.  In view of this additional interest in such identities, and
since other interesting related identities also arise via Lovelock's identities,
we give further details separately in Edgar and H\"oglund (1995).}

\medskip

\DK (1994a) also check the accuracy of
the  {\it non-vacuum}
wave equation  for the Lanczos potential  as given by Illge (1988) in
both spinor and tensor form. (But it should be noted that,  whereas
Illge considered  potentials  in algebraic Lanczos gauge but
arbitrary differential gauge of Weyl candidates, 
Dolan and Kim
(1994a) restrict their considerations to potentials in both Lanczos gauges of
Weyl curvature tensors;  so their check is only valid for potentials in 
Lanczos gauges of Weyl tensors.) 
 \DK confirm that   Illge (1988) has given
the correct {\it spinor} version of the {\it non-vacuum}
wave equation  for the Lanczos potential in Lanczos gauges. 
They also  point out  that \II has omitted a
term containing the Ricci scalar $R $ in his translation to the {\it tensor} 
version.

\medskip
An interesting more general point is that although Lanczos claimed that a 
potential for the Weyl tensor could only exist in 4-dimensional spacetimes, he
gave no rigorous arguments to support this. (However it should be appreciated
that his work was in the notation of Hodge duals and so  no generalisation
to other dimensions was possible within that formalism.) On the other hand the
only existence proofs we have, at present, are for four dimensions, given in
Bampi and Caviglia (1983) (irrespective of signature) and in
  Illge (1988) (restricted to Lorentz signature); and in the technicalities of
both proofs dimension does play a significant role. As we note in Section
2, it is  straightforward to generalise the form of (1.3) to arbitrary $n$
dimensions, but we have no firm evidence either way as to whether such a
potential exists for {\it all} Weyl tensors in any dimension $n>4$.  However, as
we shall show in Section 2, the simple homogeneous  wave equation (1.8)  exists
{\it only} in four dimensions. In   Bampi and Caviglia (1983) there is some
discussion for spaces with dimension $n>4 
$\footnote{${}^{\dag}$}{Bampi and Caviglia (1983) state, for the parallel
problem which disregards the cyclic property of the Weyl tensor, that, under
generic conditions,  a potential can exist only in spaces with  dimension $n\le
6  $;  this has lead to the careless and incorrect statement that the   Lanczos
potential for a Weyl tensor  exists only in dimensions $4\le n\le 6 $  in
Roberts (1989),
 Edgar (1994a), Edgar (1994b). Firstly, there seems to be a simple 
computational mistake in the very last step of the argument so that the 
result of  Bampi and Caviglia 
 for the parallel problem
should instead be $n\le 5  $;
secondly it does not seem obvious  
that this parallel problem 
is
equivalent to the original existence problem for the  potential of a Weyl
candidate {\it in 5
dimensions}; thirdly  a generic result for  Weyl candidates
cannot be directly applied to all Weyl curvature tensors.}; however, this is for
{\it a parallel existence  problem} (which is shown to be equivalent to the
existence problem for  potentials of Weyl candidates {\it only in four
dimensions}), and so we believe that it is still an open question  whether there
exists potentials with the  properties (1.2) and (1.4) which satisfy the 
$n-$dimensional generalisation of (1.3) for  Weyl tensors in spaces with 
dimension $n>4$.

\

 \NV (1987) have supplied some algorithms for calculating the Lanczos potential
of the Weyl conformal curvature tensor in a number of different classes of
perfect fluid spacetimes. Their method of establishing these results was simply
to substitute into the right hand side of (1.11a) certain proposed forms for the
Lanczos potential, and by direct manipulation --- a non trivial calculation ---
show that the Weyl tensor of the background spacetime is obtained. It is
emphasised that the \NR method is explicitly for {\it Weyl conformal curvature
tensors} and that they showed {\it explicitly} that their resultant 4-tensor was
such a tensor. The work in Novello and Velloso (1987) was, in general, for an
arbitrary gauge, although in some cases the form proposed for \h satisfies the
Lanczos algebraic gauge; in other cases a simple modification
$$\eqalign{L_{abc}\rightarrow L_{abc}-2g_{c[a}L^e{}_{b]e}/3
}\eqno(1.13)
$$ always yields a Lanczos potential which satisfies the algebraic
Lanczos gauge. When 
 their algorithms were applied to particular metrics, it was often found  that
the Lanczos differential gauge was also satisfied. However, as pointed out in
 Novello and Velloso (1987), there is no compelling
argument, in general,  to prefer the Lanczos differential gauge.

 \medskip

In the second of their   recent papers,  Dolan and Kim (1994b)
have  used and developed some of the results of Novello and
Velloso; we feel that some points in this paper require
 further comment, and we deal with these points
later in this paper. 

\

In the next section we present, for the first time, the
most general form of the second order  differential equation for the
 3-potential of a Weyl candidate which satisfies a Bianchi-type equation 
--- 
 in  $n$-dimensional spaces with  arbitrary signature, and with
potential  in arbitrary gauge --- {\it for such spaces  for which such a
potential exists}; from this we can deduce, also for the first time, the form
in arbitrary four dimensional spaces with arbitrary signature and with
potential  in arbitrary gauge.  (Since these calculations have been carried out
by computer  we hope it will end the need for further checks.) We show
explicitly that for Lanczos potentials of Weyl tensors the  non-linear
complications  disappear and the awkward second order terms cancel {\it
only} in four dimensional spaces (irrespective of signature);  further a
remarkably simple form of this four dimensional equation seems to occur   only  
in the Lanczos gauges, and the Lanczos-Illge wave equation (1.8) is the vacuum
version of this for Lorentz signature. We also confirm that the non-vacuum
tensor form  for the wave equation of the Lanczos potential in Lanczos gauges
given in  Dolan and Kim (1994a) is correct, and agree with them that Illge's
version has a  missing term  involving the Ricci scalar;  however, contrary to
what is claimed in   Dolan and Kim (1994a), we have found that this is the {\it
only} error in the tensor version in Illge (1988). (We have also checked the
more general tensor version, in arbitrary differential gauge, as given by Illge
(1988).)

In Section 3 we deal directly with some points from the second of the papers by
Dolan and Kim (1994b).  In that paper they suggest three {\it possible} classes
of Lanczos potentials (for Weyl tensors) in Lanczos gauges, and propose that
`when these are verified for each spacetime to which they are applied', then
they will be genuine Lanczos potentials in Lanczos gauges. However, one of these
classes is simply a subclass of a class already shown in Novello and Velloso
(1987) to always give Lanczos potentials in Lanczos gauges; so no further
explicit verification is necessary. A second of these classes is precisely the
vacuum subclass of a class already shown in Novello and Velloso (1987) to always
give Lanczos potentials --- although not necessarily in Lanczos gauges; so only
the gauge conditions need to be explicitly verified. The remaining class
suggested in Dolan and Kim (1994a) is a new proposal, and so for this class
explicit verification of both the defining equation (1.11a) for Weyl tensors,
and for the Lanczos gauges (1.4) is {\it essential} --- but this is not given in
Dolan and Kim (1994b) for the application to the Bondi metric.  In Section 5 we
explicitly confirm the validity of this potential in general for the class
proposed by \DK, and in particular for the vacuum Bondi metric.

In Section 3, we also point out an unfortunate structural error in the
`Weyl-Lanczos' relations given by Dolan and Kim (1994b). Their spin coefficient
version of (1.11) is inconsistent with their definitions for the tetrad/dyad
components of the Lanczos potential components. To maintain internal
consistency, as well as agreement with the earlier work in Zund (1975) and in
Ares de Parga et al. (1989), we propose that the {\it definitions of the Lanczos
potential components} (both tetrad and dyad version), in Dolan and Kim (1994b),
be adjusted by a factor of $(-1)$.  A similar change is also required in the two
applications which use these definitions.

In Section 4 we establish the perhaps surprising result that in a background
four dimensional vacuum spacetime, the homogeneous wave equation (1.8a) for a
3-tensor $L_{abc} $ in the Lanczos gauges, is a {\it sufficient} condition ---
in general --- for $L_{abc} $ to be a Lanczos potential (up to a constant
factor) of the Weyl curvature tensor of this background spacetime. In Section 5
we use this result to confirm, in a very concise manner, the validity of the new
`possible' Lanczos potential proposed by \DK (1994b) --- in most vacuum metrics
with the appropriate symmetries; this includes their proposed new Lanczos
potential for the vacuum Bondi metric. We also investigate whether this
potential is valid for any non-vacuum Bondi spaces, but unfortunately find that
the only physically interesting generalisation seems to be to vacuum spaces with
non-vanishing cosmological term.

\

Although we shall not deal directly with the results in this paper we draw
attention to other recent related work: Hammon and Norris (1993) consider the
Lanczos potential, and in particular the question of gauge, in a still more
general geometric setting; Torres del Castillo (1995) has obtained Lanczos
potentials for a large class of spacetimes, including the Kerr metric, by spinor
means; Ares de Parga G., et al (1989) and L\'opez-Bonilla, J.L. et al (1993)
have obtained expressions for the Lanczos potentials of some special spacetimes,
in the spin coefficient formalism of Newman and Penrose (1962).

\

\vfill\eject

\beginsection 2. The general second order 
differential equation for the Lanczos
potential.

Although we have no evidence to believe that in $n 
$ dimensions {\it every} 
Weyl candidate $W_{abcd} $ satisfying (1.1) can be 
given locally in terms of a
3-tensor potential $L_{abc}$ satisfying (1.2), we can of course 
define a class of Weyl
candidates by the $n $-dimensional generalisation of 
(1.3),
$$\eqalign{ W&_{abcd}  =  \  2 L_{ab[c;d]}
+2L_{cd[a;b]}  -{2\over (n-2)} 
g_{a[c}(L_{|b|}{}^e{}_{d];e}-L_{|b|}{}^e{}_{e;d]}+
L_{d]}{}^e{}_{b;e}-L_{d]}{}^e{}_{e;b}) \cr & 
+{2\over (n-2)}g_{b[c}(
L_{|a|}{}^e{}_{d];e}-L_{|a|}{}^e{}_{e;d]}+
L_{d]}{}^e{}_{a;e}-L_{d]}{}^e{}_{e;a})  +{8\over 
(n-2)(n-1)}
g_{a[c}g_{d]b}L^{ef}{}_{e;f}}  \eqno(2.1) $$
For $n=4 $ this expression coincides with (1.3) 
and is valid for all Weyl
candidates; but for $n>4 $ we do not know whether 
the class so defined is the
set of {\it all} Weyl candidates.

It is  assumed that the Weyl candidate satisfies a 
Bianchi-like equation
$$W_{abcd;}{}^d=J_{abc}
\eqno(2.2)
$$
When we substitute (2.1) in (2.2) we obtain a 
second order differential 
equation
for the potential $L_{abc} $  given by
$$\eqalign{
&
  \nabla^2 L_{abc}
  +\frac{2(n-4)}{n-2}L_{[a}{}^{d}{}_{|c;d|b]}
  -\frac{3}{(n-2)(n-1)}RL_{abc}
\cr&
  -2L_{[b}{}^{ed}C_{a]dec}
  +\frac{1}{2}C_{deab}L^{de}{}_{c}
  -\frac{4}{n-2}g_{c[a}C_{b]fed}L^{fed}
\cr&
  +\frac{2(n-4)}{n-2}R_{d[a}L_{b]}{}^{d}{}_{c}
  -2R_{d[a}L_{b]c}{}^{d}
  +\frac{n}{n-2}R_{cd}L_{ab}{}^{d}
  +\frac{2n}{(n-2)^2}g_{c[a}L_{b]ed}R^{ed}
\cr&
  +G_{abc}
  \qquad\qquad\qquad\qquad=\ J_{abc}
}\eqno{(2.3)}$$
where
$$\eqalign{G_{abc}&=\frac{1}{n-2}\Bigl(
  2g_{c[a}\nabla^2 L_{b]}{}^{e}{}_{e}
  +C_{abcd}L^{de}{}_{e}
\cr&
  +\frac{2(n-5)}{n-1}L^{ed}{}_{d;e[b}g_{a]c}
  -2g_{c[a}L_{b]}{}^{ed}{}_{;de}
\cr&
  +2L_{[a}{}^{e}{}_{|e;c|b]}
  -2(n-3)L_{[a|c}{}^{d}{}_{;d|b]}
  -(n-2)L_{ab}{}^{d}{}_{;dc}
\cr&
  +\frac{2(n-1)}{n-2}R_{c[a}L_{b]d}{}^{d}
  -\frac{2(n-5)}{n-2}g_{c[a}R_{b]e}L^{ed}{}_{d}
  -\frac{2(n+1)}{(n-2)(n-1)}g_{c[a}L_{b]d}{}^{d}R
\Bigr)
}\eqno(2.4)$$
 This equation is valid in arbitrary
gauge,
 in arbitrary $n$ dimensions with arbitrary 
signature; $G_{abc} $ vanishes
when both Lanczos gauges are applied.

\medskip

We now look at some special cases of this 
equation, and compare them with
existing equations in the literature. We remember 
that although Illge (1988) gave
a tensor version of this equation, this was 
deduced from his
spinor version, and so  was strictly only valid in 
4-dimensional spacetimes
(with Lorentz signature); also the 3-potential of the 
Weyl candidate was in Lanczos
algebraic gauge.  \DK (1994a), on the otherhand, give both 
a spinor and an independent
4-dimensional tensor derivation, and so their 
tensor version is  valid for any
 signature in four dimensions; however, their 
considerations were explicitly restricted
to Lanczos potentials in Lanczos gauges of Weyl 
curvature tensors.   

\ 

{\bf (i) Four dimensions.} 

In 4-dimensional spaces with arbitrary gauges 
Edgar and H\"oglund 
(1995) show that the 
identity (1.10) generalises to,
$$
\eqalign{
 2L^{def}g_{[a|c|}C_{b]def}
-2L_{[a}{}^{de}C_{b]edc}-{1\over2} L^{de}{}_c 
C_{deab}
-{1\over2}C_{abcd}L^{de}{}_e
=0}
\eqno(2.5)
$$
Hence in (2.3), in four dimensions, the product terms of $L_{abc} $ with
$C_{abcd} $ disappear and so do the awkward second derivative terms, because
they involve a factor of $(n-4) $, so that (2.3) becomes
$$
\eqalign{
&
  \nabla^2 L_{abc}
   -2R_{d[a}L_{b]c}{}^{d}
  +2R_{cd}L_{ab}{}^{d}
  +2g_{c[a}L_{b]ed}R^{ed}
-\frac{1}{2}R L_{abc}+G_{abc}
  = J_{abc}
}\eqno{(2.6)}
$$
where
$$\eqalign{G_{abc}&=
  g_{c[a}\nabla^2 L_{b]}{}^{e}{}_{e}
  -\frac{1}{3}L^{ed}{}_{d;e[b}g_{a]c}
  -g_{c[a}L_{b]}{}^{ed}{}_{;de}
  +L_{[a}{}^{e}{}_{|e;c|b]}
  -L_{[a|c}{}^{d}{}_{;d|b]}
  -L_{ab}{}^{d}{}_{;dc}
\cr&
  +\frac{3}{2}R_{c[a}L_{b]d}{}^{d}
  +\frac{1}{2}g_{c[a}R_{b]e}L^{ed}{}_{d}
  -\frac{5}{6}g_{c[a}L_{b]d}{}^{d}R
}\eqno(2.7)
$$
 So it has
now been shown that {\it all} product terms of $L_{abc} $ with
$C_{abcd} $ 
disappear in {\it any} gauge of the
Lanczos potential; we have also shown, 
 contrary to what was implied by Dolan  and
Kim (1994a), that these product terms disappear 
for {\it all} signatures of
4-dimensional spaces.

It may  be possible to
eliminate $G_{abc} $ (which contains all the  terms arising from arbitrary 
gauge) by  gauge choices other
than the two Lanczos gauges, but there are no 
obvious simple alternatives.

\

{\bf (ii) Four dimensions,  Lanczos algebraic 
gauge.}

Equation (2.6) is unchanged 
with (2.7) becoming
$$
\eqalign{G_{abc}&=
  -g_{c[a}L_{b]}{}^{ed}{}_{;de}
  -L_{[a|c}{}^{d}{}_{;d|b]}
  -L_{ab}{}^{d}{}_{;dc}
}\eqno(2.8)
$$
This  agrees  with  the tensor version (in Lorentz 
signature) given by
Illge (1988) --- subject to the addition of the 
Ricci
scalar term, and the different  conventions.

\

{\bf (iii) Four dimensions, Lanczos gauges.} 

Equation (2.6)  becomes 
$$\eqalign{
&
  \nabla^2 L_{abc}
  -2R_{d[a}L_{b]c}{}^{d}
  +2R_{cd}L_{ab}{}^{d}
  +2g_{c[a}L_{b]ed}R^{ed}
  -\frac{1}{2}RL_{abc}
    =J_{abc}
}\eqno{(2.9)}
$$

This  of course also agrees with Illge (1988) --- 
subject to the addition of the Ricci
scalar term;
 and when the Weyl candidate $W_{abcd}$ is
specialised to the Weyl curvature tensor 
$C_{abcd}$, there is also agreement with
the version given by Dolan and Kim 
(1994a).\footnote {$^{\dag} $}{The additional 
 disagreement claimed by  Dolan and Kim (1994a), 
between  
Illge's tensor equation and their own, is only 
{\it
apparent}; it disappears when we take into account 
the
  different conventions for the Riemann, Ricci and 
Lanczos tensors between  Dolan
and Kim (1994a) and  Illge (1988).  These are also 
the conclusions of Illge 
(1995).}

\

{\bf (iv) 
Four dimensions, vacuum, Lanczos gauges, Weyl 
curvature tensor.}

We replace the Weyl candidate $W_{abcd} $ with the 
Weyl curvature 
tensor $C_{abcd} $, and replace (2.2) with the 
Bianchi equations so that $J_{abc}=0 $.  
Equation (2.9) becomes   
$$\eqalign{\nabla^2 L_{abc}=0
}\eqno(2.10)$$
When we specialise to Lorentz signature we recover 
the Lanczos-Illge wave
equation (1.8b).

\

{\bf (v)  $n $-dimensions ($n >4$).} 

We have noted in (i) that the expression giving 
rise to the
product terms of $L_{abc} $ with $C_{abcd} $
 $$\eqalign{
  -2L_{[b}{}^{ed}C_{a]dec}
  +\frac{1}{2}C_{deab}L^{de}{}_{c}
  -\frac{4}{n-2}g_{c[a}C_{b]fed}L^{fed}
  +\frac{1}{n-2}C_{abcd}L^{de}{}_{e}
}\eqno(2.11)$$
has been shown to be identically zero in four 
dimensions. It has also been
shown by H\"oglund (1995) (for Lanczos gauges) and 
Edgar and H\"oglund (1995) 
(for arbitrary gauges) that (2.11) cannot be 
identically zero 
for dimensions $n>4 $.\footnote{$^{\dag} $}{ It is 
interesting to note that a similar situation
arises in the case of the Penrose wave equation 
for the Weyl tensor; the product
terms in this case  --- involving the Weyl and 
Ricci tensors --- also only
disappear in $n=4 $ dimensions, as shown by 
Andersson and Edgar (1995).} We also
note that there are  terms in (2.3) involving 
second order derivatives and
factors of $(n-4)$ which cannot vanish for 
dimensions $n>4 $. Lanczos gauge
choices do not cause any further simplification of 
the 
product terms of $L_{abc} $ with $C_{abcd} $, nor of
the two terms involving second order  derivatives, 
and it is difficult to see
any other gauge choices which would be any more 
successful.
It is clear that in dimensions $n>4 $ the form of 
the second order differential
equation for the 3-potential of a Weyl candidate 
has a much more complicated
structure than in the 4-dimensional case; and it 
is remarkable that it is
{\it only} in four dimensions that such major 
simplifications occur.

\

{\bf (vi) $n$-dimensions, vacuum, Lanczos gauges, 
Weyl curvature tensor.} 

We replace the Weyl candidate $W_{abcd} $ with the 
Weyl curvature 
tensor $C_{abcd} $, and replace (2.2) with the 
Bianchi equations. 
With the additional vacuum and gauge 
simplifications,  (2.3) 
simplifies to
$$
\eqalign{
&
  \nabla^2 L_{abc}
  +\frac{2(n-4)}{n-2}L_{[a}{}^{d}{}_{|c;d|b]}
\cr&
  -2L_{[b}{}^{ed}C_{a]dec}
  +\frac{1}{2}C_{deab}L^{de}{}_{c}
  -\frac{4}{n-2}g_{c[a}C_{b]fed}L^{fed}
  =0
}
\eqno(2.12)
$$

However, this equation is even more complicated 
than its appearance
suggests; we also have to use (1.11a) to 
substitute for the  Weyl tensor \c,
and this results in  complicated non-linear terms 
involving
products of the Lanczos potential with its first 
derivatives.  
\smallskip
Once again, it is remarkable that this very 
complicated non-linearity 
for Lanczos potentials of Weyl tensors
disappears 
{\it
only}  in four dimensions. 

\ 

These calculations and results in this Section are presented in more detail in
H\"oglund (1995); they were derived with the computing techniques described
there, and checked with {\it MathTensor}, (Parker and Christensen (1991).

\vfill\eject

\beginsection 3. Some comments on  Dolan and Kim (1994b).

The purpose of the second paper by \DK was to propose some possible Lanczos
potentials in Lanczos gauges for Weyl tensors in some particular spacetimes, and
to test these potentials in the Lanczos-Illge vacuum wave equation (1.8). They
made use of some results by \NV (1987) who have constructed Lanczos
potentials for Weyl tensors in a number of different  classes of
perfect fluid spacetimes. We emphasise that the work by \NV involved
lengthy calculations whereby they showed directly that their proposed
Lanczos potentials gave rise to genuine Weyl conformal curvature tensors (with
the required differential as well as algebraic properties.)

As noted in the Introduction the choices (1.4) are optional gauge
choices. However, it is emphasised that in Dolan and Kim (1994b) these Lanczos
gauges are always implied, and so it should be understood that when \DK talk
about a `Lanczos potential' they are by definition requiring that it is a
potential of the Weyl curvature tensor and satisfies both Lanczos gauges. In
fact, they construct, where necessary, their potentials in the form (1.13) to
guarantee that the algebraic Lanczos gauge is satisfied automatically; no such
simple adjustment is possible for the differential gauge condition, so in Dolan
and Kim (1994b) the condition (1.4b) is always understood to be supplementing
the defining equation (1.11). Obviously, a change of algebraic or differential
gauge has no effect on the defining equation (1.11). To avoid misunderstandings
in this present paper we shall specifically state when the Lanczos differential
gauge is also being required.

 In  
Novello and Velloso (1987) a
number of different constructions for Lanczos potentials of Weyl tensors are
proposed; when these are applied to a variety of spacetimes, in some cases the
Lanczos potential turns out to satisfy the Lanczos algebraic and/or differential
gauge. As they pointed out  the simple modification (1.13) will always guarantee
us a Lanczos potential in the Lanczos algebraic gauge. If we also want to know
whether  the Lanczos differential gauge is satisfied, then we simply test the
Lanczos differential gauge condition (1.4b) directly. In the case of the
different forms of the Lanczos potentials proposed by  Novello and Velloso
(1987), then any which also satisfy (1.4b) will obviously automatically be
Lanczos potentials in the Lanczos differential gauge.

With the above  in mind, we now look in detail at some points 
from  Dolan and Kim (1994b): 
\medskip
{\bf 3.1. Possible Lanczos Potentials in the Lanczos  Gauges.}

In Section 2 of Dolan and Kim (1994b) three classes of `possible Lanczos
potentials [in Lanczos gauges]' are proposed for Weyl tensors; we consider each
of these in detail, and then the respective applications.

\smallskip

 {\bf (i)} The first of their classes is for vacuum spacetimes admitting a
hypersurface orthogonal timelike Killing vector $\xi $ --- the timelike version
in case (a) of Section 2 in Dolan and Kim (1994b) --- and as shown there, this
implies that the associated unit vector is hypersurface orthogonal, shear-free
and expansion-less; so this is just a subclass of the perfect fluid spacetimes
containing a hypersurface orthogonal, shear-free timelike unit vector considered
in Lemma 2 by Novello and Velloso (1987). \DK propose exactly the same form of
potential as in Novello and Velloso (1987), and since it was verified there,
{\it for the whole class}, that such a potential is a genuine Lanczos potential
in both Lanczos gauges, then clearly the subclass proposed in Dolan and Kim
(1994b) also satisfies these conditions. So therefore there is no need for any
further verification. We emphasise this point because in Dolan and Kim (1994b)
it is stated \footnote{${}^{\dag}$}{In any such quotation we replace the
equation numbers in Dolan and Kim (1994b) with the equivalent equation numbers
in this paper.}  `the [possible Lanczos potential] satisfies (1.2) and (1.4a),
and if in each application we verify (1.11a) and (1.4b) then the [possible
Lanczos potential] is a Lanczos potential tensor [in Lanczos gauges] for the
space-time'; clearly this additional individual verification is unnecessary.

\smallskip 

{\bf (ii)} The third of their classes is for
vacuum spacetimes admitting
 a unit hypersurface orthogonal timelike geodesic field with tangent vector $v $
--- case (b) in Section 2 in  Dolan and Kim (1994b). This is precisely the
vacuum subclass of a class considered by \NV in Lemma 3(ii), and
\DK propose exactly the same potential. In  Novello and Velloso (1987)  it was
shown directly that this proposal is a genuine  Lanczos potential,  but it was
{\it not} shown there that this potential satisfies either
of the Lanczos gauges.   So therefore there is no need
for any further verification of the defining equation (1.11a), but the two
Lanczos gauges (1.4a,b) need to be checked explicitly. We emphasise this point
because  in
 Dolan and Kim (1994b) it is again stated    
 `the  [possible Lanczos potential] will satisfy (1.2) and (1.4a) and when we
verify (1.11a) and (1.4b) then it will be a Lanczos
potential tensor [in Lanczos gauges] for the space-time'; but as we have just
noted (1.4a) has not been checked explicitly, whereas (1.11a) has been
confirmed explicitly in  Novello and Velloso (1987). However, in  Appendix I we
easily show that the Lanczos algebraic gauge (1.4a) is always satisfied; but in
general, the Lanczos differential gauge (1.4b) is not. Therefore if we wish to
have a Lanczos potential in the Lanczos gauges for this class, we will need to
check {\it only} (1.4b) individually for each application.

\smallskip  

{\bf (iii)} Their second class of `possible Lanczos potentials [in Lanczos
gauges]' is for vacuum spacetimes admitting a hypersurface orthogonal spacelike
Killing vector $\Xi $ --- the spacelike version in case (a) of Section 2 in
Dolan and Kim (1994b); this is the spacelike counterpart of the first proposal
discussed in (i) above, and the proposed new form for the Lanczos potential
$L^{(2)}_{abc} $ is analogous to the form of the Lanczos potential, discussed
above in (i).  Since this is a new proposal it is then {\it absolutely
necessary} --- either in a generic manner like in Novello and Velloso (1987), or
for each individual application for which this form of potential is chosen ---
to verify explicitly first of all that (1.11a) is satisfied, and secondly that
(1.4b) is satisfied. (The form of the proposed potential guarantees that (1.2)
and (1.4a) are automatically satisfied.)  However, in Dolan and Kim (1994b),
these equations are {\it not} verified in a generic manner; and although it is
stated again `the [possible Lanczos potential] satisfies (1.2) and (1.4a) and if
in each application we verify (1.11a) and (1.4b) then the [possible Lanczos
potential] is a Lanczos potential tensor [in Lanczos gauges] for the
space-time', {\it the explicit verification for the application to the Bondi
metric is {\it not} given.}

We emphasise that the point being made here is not that we believe the form
$L^{(2)}_{abc} $ for this class is wrong, but that there is no explicit proof or
statement in Dolan and Kim (1994b) as to {\it why} it has been concluded that
$L^{(2)}_{abc} $ is a genuine Lanczos potential of the Weyl tensor of the Bondi
spacetime (although an implicit confirmation for the Schwarzschild application
is stated).

Since we believe that  it has still to be explicitly confirmed that
$L^{(2)}_{abc} $ given for the Bondi metric is indeed a Lanczos potential for
the Weyl tensor in the Lanczos gauges, we will discuss this point further  and
give explicit verification  in  Section 5.

\smallskip

{\bf (iv)} In the remainder of Section 2 of Dolan and Kim (1994b) the three
different classes of Lanczos potential in Lanczos gauges are applied to some
specific spacetimes.  The application (Schwarzschild) chosen to illustrate the
class described in (i) above is (as \DK note) a spacetime for which an explicit
Lanczos potential in Lanczos gauges has already been verified in Novello and
Velloso (1987), but we emphasise that this verification was made in Novello and
Velloso (1987) for the whole class --- they did not have to verify this case
individually.

The application (Kasner) chosen to illustrate the class described in (ii) above
is (as \DK note) a spacetime for which an explicit Lanczos potential in Lanczos
gauges has been verified in Novello and Velloso (1987), but we emphasise that
its status as a Lanczos potential was verified as part of the whole class ---
they did not have to verify the Lanczos gauge individually.

For the new `possible Lanczos potential', described in (iii) above, applications
are given to both the Schwarzschild and Bondi metrics.  There seems some
ambiguity in Dolan and Kim (1994b) as to when such a `possible Lanczos
potential' is considered a genuine Lanczos potential. For instance, in the
Schwarzschild application in Section 2, a potential of this class $L^{(2)}_{abc}
$, is proposed as a `possible Lanczos potential', and at this point no
justification is given that it is indeed a Lanczos potential of the Weyl tensor
in this particular spacetime i.e. whether it satisfies (1.11) --- although a
comment to this effect is made later, in Section 3. On the other hand, in their
next application, also in Section 2, on the Bondi spacetime, it is immediately
stated, without any discussion, that $L^{(2)}_{abc} $ (defined as before) is a
Lanczos potential; we feel that this conclusion is premature, since the promised
verification in the application to the Bondi metric has not been carried out
explicitly, neither for the defining equation (1.11a) nor for the Lanczos
differential gauge condition (1.4b).

\medskip  

{\bf 3.2. The homogeneous wave equation for the Lanczos potential in
Lanczos gauges,  in vacuum spacetimes.}

As noted in the Introduction the particularly simple form (1.8) of the wave
equation for $L_{abc} $ occurs in vacuum spacetimes, in the Lanczos gauges.  \DK
set out to provide some `solutions of this wave equation'. Their procedure is
simply to choose some special vacuum spacetime, write down --- using examples
from each of the three algorithms discussed above --- a potential in the Lanczos
gauge for that spacetime, and then test explicitly the Lanczos-Illge wave
equation (1.8) for that particular Lanczos potential in that particular
spacetime. However, for known Lanczos potentials in Lanczos gauges, in vacuum,
there is clearly no need to check the wave equation explicitly in each
individual case --- it follows automatically; calculations for individual
spacetimes are just special cases of the general case.

On the other hand, for new `possible Lanczos potentials', the simple wave
equation is not guaranteed.  However, in Appendix I, when we investigate --- as
a whole --- the new class of possible Lanczos potentials referred to in (iii)
above, we discover that all members of this class do in fact satisfy the simple
wave equation (1.8). So again there is no need to consider individual cases in
this class separately as is done in Dolan and Kim (1994b). In addition, this
result becomes very significant when considered alongside our result in Section
4.

\medskip

{\bf 3.3. The Weyl-Lanczos equations.}

Our discussion on this topic relates to both Dolan and Kim (1994b) and to other
papers; but we especially wish to draw attention to a structural mistake in the
version of these equations in Dolan and Kim (1994b).  \smallskip {\bf (i)} The
first translations of the Lanczos tensor into spinor notation seems to be due to
Maher and Zund (1968), and the first attempt to write down a direct version of
(1.11) using NP spin-coefficients, --- the {\it five} complex Weyl-Lanczos
equations --- also seems first to have been made in Maher and Zund (1968);
however, the notation is unwieldy, and there are a number of errors in that
paper. A reliable spinor presentation is given by Taub (1975), and an improved
presentation of the Weyl-Lanczos equations, in NP formalism, appeared in Zund
(1975) (again with errors, but fewer), together with the NP formalism version
(three complex equations) of the Lanczos differential gauge (1.4b). A more
reliable version of the Weyl-Lanczos equations is given in Ares de Parga et
al. (1989) where there seems to be just one rather obvious misprint --- in the
expression for $\Psi_2 $ the coefficient of $\Omega _3 $ should be $\sigma $
rather than $\alpha $; unfortunately, when some of the same group of authors
attempt to quote these equations in a later paper by L\'opez-Bonilla et al.
(1993) there are a number of misprints.

It is emphasised that in  both  Zund (1975)  and in Ares de Parga 
et al. (1989) the equations were given in arbitrary
 differential gauge. 
 Although   Zund (1975)  uses spinor dyads and Ares de Parga  et al.(1989) uses
tensor tetrads they essentially agree  on their definitions and notation for the
eight complex independent  components of the Lanczos
potential, $\Omega_0, ...,\Omega_7 $ --- up to a factor of $2 $ on all
components. \footnote{${}^{\dag}$}{ One has to take care since these two papers
use different definitions for the Weyl tensor. Also Ares de Parga 
et al.(1989) use the conventions of Kramer et al.(1980) for the ordering,
numbering and labelling of their tetrad vectors and NP spin coefficients ---
but with one confusing variation; instead of labelling the tetrad vectors
$\{m,\bar m,l,k\}$ as in Kramer et al.(1980), they label them as $\{m,\bar
m,l,n\}$.}

\smallskip
{\bf (ii)} \DK (1994b) propose a set of equations --- (3.5)--(3.12)  --- which
they also call the `Weyl-Lanczos equations'. These {\it eight} complex equations
are a  direct  NP spin coefficient version of  
$$\Psi_{ABCD}=2\nabla_{A}{}^{\dot A} L_{BCD} $$ (which is (1.11b) combined with
the Lanczos differential gauge (1.6)) and so consists of the five original
Weyl-Lanczos equations --- as given in Zund (1975), Ares de Parga et al. (1989)
--- {\it combined with the three equations which are the spin coefficient
version of the Lanczos differential gauge} (1.6). Of course, it is
straightforward to separate this set into its two parts. However, there is no
mention of the Lanczos differential gauge condition in the section in Dolan and
Kim (1994b) on the spin coefficient version; to avoid any misunderstanding this
should be noted, so that the more complicated structure of this set of equations
be understood.  Since this set of NP equations in \DK (1994b) include {\it both}
the defining equations for the Lanczos potential and the Lanczos differential
gauge, they alone need to be checked explicitly for new proposed Lanczos
potentials in the Lanczos differential gauge.

\smallskip

{\bf (iii)} When we compare the definitions (in both vector and spinor notation)
for the eight complex independent components of the Lanczos potential, $L_0,
...,L_7 $ in Dolan and Kim (1994b) with the set of Weyl-Lanczos equations given
there --- as differential equations for $L_0, ...,L_7 $ --- we see that there is
a systematic inconsistency in sign.\footnote{${}^{\dag}$}{We believe that this
is due to a mistake in sign when lowering an index (or a departure from the
conventions of Penrose and Rindler(1984)) going from equation (3.2) to (3.3) in
Dolan and Kim (1994b).} In fact, the definitions of the components of the
Lanczos potential, $L_0, ...,L_7 $ in Dolan and Kim (1994b) are the negative of
the definitions of $\Omega_0, ...,\Omega_7 $ in Ares de Parga et al. (1989) and
Zund (1975), although the form of the equations in Dolan and Kim (1994b)
essentially agrees with the (Lanczos differential gauge version) of the
equations in Ares de Parga et al.(1989) and Zund (1975). So we propose that,
\medskip 1. {\it The definitions (in both tensor and spinor notation) for the
eight complex independent components of the Lanczos potential, $L_0, ...,L_7 $
in Dolan and Kim (1994b) all be adjusted by a factor of $(-1) $, but their
Weyl-Lanczos equations remain unchanged}.  \smallskip 2. {\it The values of all
explicit Lanczos potential tetrad/dyad components be adjusted by a factor of
$(-1) $; in particular this applies to the Section 3 examples for the
Schwarzschild and Bondi spacetimes.}  \medskip These changes will not only
correct Dolan and Kim (1994b), but will be consistent with the definitions and
equations in Ares de Parga et al. (1989) (and with Zund (1975), subject to the
factor of 2, and the misprints there). We shall assume this correction for the
rest of this paper, and correct their applications in (iv) below, and in Section
5.

\smallskip

{\bf (iv)}  The application for the Schwarzschild example --- in Section 3 of
Dolan and Kim (1994b) --- makes explicit use of their `Weyl-Lanczos equations',
and rather surprisingly --- in view of the fact that we now know that these
equations  are inconsistent with the definitions of the Lanczos potential
components as proposed in  Dolan and Kim (1994b) --- it is claimed that their 
Weyl-Lanczos equations are satisfied. It is certainly true that when the given
Lanczos potential components are substituted into their given Weyl-Lanczos
equations --- for both of the proposed Lanczos potentials
$L_{abc}^{(1)}, L_{abc}^{(2)} $ --- we do find the only non-zero component of
the Weyl tensor is $\Psi_2=m/r^3 $, as  stated in
 Dolan and Kim (1994b). However, in the notation and conventions of 
Penrose and Rindler (1984), the
value should be    
$\Psi_2=-m/r^3
$. (This value  can easily be confirmed by substituting the
spin coefficients and differential operators as given by \DK (1994b) into the
relevant NP equation in Penrose and Rindler (1984)). Hence we find that the
calculations in this application {\it do not verify} the Weyl-Lanczos equations
for the definitions of  the Lanczos potential components as proposed in  Dolan
and Kim (1994b); this is of course as we would now expect, and it is clear that
the adjustment by a factor of $(-1) $, proposed above,  on all  the Lanczos
potential components, will ensure agreement.

 This corrects the confirmation by direct calculations  that
$L^{(2)}_{abc} $ is a genuine Lanczos tensor in the Lanczos gauge for the 
Schwarzschild metric.

\medskip

{\bf 3.4. The Lanczos-Illge wave equation as a condition for a Lanczos
potential.}

Finally we would point out that one reason for stressing these points in detail
is the fear --- since \DK do not in most cases confirm explicitly the defining
equation (1.11a) and the Lanczos gauge condition (1.4b), but rather concentrate
on the homogeneous Lanczos wave equation (1.8) --- that some readers may
mistakenly conclude that the homogeneous wave equation is a sufficient (as well
as a necessary) condition for a Lanczos potential of the background Weyl tensor
in the Lanczos gauges in vacuum spacetimes. There is of course no reason, at
this stage, to suspect this; on the other hand we shall show in the next section
that this is often the case.

\ 

\

\vfill\eject

\beginsection 4. Lanczos-Illge wave equation as a sufficient condition for 
Lanczos potential.

If the   simple homogeneous wave equation (1.8) for the Lanczos potential \h
of the Weyl curvature tensor is going to play a significant role in general
relativity, or even in differential geometry,  then it is essential to be able
to distinguish between generic potentials for any {\it Weyl candidate}
satisfying Bianchi-like equations, and the Lanczos potentials for  {\it Weyl
curvature tensors/spinors}.

\medskip As pointed out in Section 1, Illge (1988) has shown that any Weyl
candidate (fully symmetric) spinor $W_{ABCD} $, can always be given locally in
terms of a four-index spinor potential $L_{ABC\dot A} $ (which is completely
symmetric in its three undotted indices) as $$ \SU W_{ABCD}=2\nabla_{D}{}^{\dot
A} L_{ABC\dot A} \eqno(4.1)$$ in the Lanczos differential gauge
$$
\nabla^{A \dot A}L_{ABC\dot A}=0
 \eqno(4.2)$$

Further, if the Weyl
candidate   satisfies 
$$
\nabla^{D\dot A}W_{ABCD}=0\eqno(4.3)
$$
it necessarily follows  that its potential satisfies
$$\square L_{ABC\dot A} = 0\eqno(4.4)$$
in vacuum.

\medskip 
The Lanczos potential of the  Weyl conformal 
 spinor $\Psi_{ABCD} $,
  by virtue of the Bianchi equations,  
 also necessarily satisfies (4.4) in vacuum, in the Lanczos gauge. At first
sight it would appear unlikely that  (4.4) is also a {\it sufficient} condition
for a potential to be the  Lanczos potential of the {\it Weyl conformal spinor}
of the background vacuum spacetime, but we shall show that this is essentially
true.

Although we have no reason to expect an arbitrary Weyl candidate $W_{ABCD} $
satisfying (4.3) in a vacuum spacetime to be the Weyl spinor of the background
vacuum spacetime, nor indeed of any spacetime, of course one of these Weyl
candidates is the Weyl spinor of the background spacetime.  However, Bell and
Szekeres (1972) have shown --- in a given background vacuum spacetime of
sufficient generality --- that (4.3) has a unique solution up to a constant
factor. Since the Weyl spinor $\Psi_{ABCD} $ of the background spacetime also
satisfies (4.3) then --- in general circumstances --- $$ \Psi_{ABCD}= k
W_{ABCD}\qquad\qquad\hbox{where $k $ is constant}\eqno(4.5) $$ Therefore, if in
a vacuum spacetime, we construct a Weyl candidate by (4.1) from a potential
satisfying (4.2) and (4.4) then, in general, that Weyl candidate is actually the
Weyl conformal spinor of the background vacuum spacetime (up to a constant
factor).

Summing up: {\it 
In a vacuum spacetime of sufficient generality, equations (4.2) and (4.4) are 
sufficient conditions that the potential $kL_{ABC\dot A} $ be a Lanczos
potential for the Weyl curvature spinor given by  $\SU \Psi_{ABCD} =k\nabla_D{}^
{\dot B} L_{ABC\dot B} $ of   the background vacuum spacetime, where $k $ is a
constant factor to be determined. }

\smallskip
There are two distinct special situations when (4.3) fails to have a unique
solution (up to a constant factor) and hence our result does not hold:

(a) the background vacuum spacetime  is algebraically special; 

(b) the background vacuum spacetime is algebraically general but satisfies a
very restrictive condition, given explicitly in Bell and Szekeres (1972). For
this class, the solutions to (4.3) are linear combinations of, at most, two
independent solutions. Therefore it is possible, for this special class, that a
solution of (4.4) would give a Weyl candidate which is completely independent of
the Weyl spinor.  \medskip It should be noted that it has been shown explicitly
in Bell and Szekeres (1972) that there is a nontrivial additional term
introduced into (4.5) for the exceptional case (a), and it has also been shown
that case (b) is not empty by the construction of a counterexample to (4.5).
This means that it is not just that we have been unable to prove our result for
the Lanczos potential in these two exceptional cases, but that in these cases
our result cannot be proven.

\

Finally, we emphasise that not only are the results in this section valid only
in four dimensions but --- since the uniqueness result of \BS has been derived
using spinors --- the results are only valid in spacetimes, with Lorentz
signature; it remains to be seen whether these results can be generalised to
other signatures and dimension.

\vfill\eject 

{\bf 5. Verification that the new proposal in  Dolan and Kim (1994b) gives a
Lanczos potential for certain vacuum  spacetimes.} 

We consider explicitly  the proposal in  Dolan and Kim (1994b) for a
$`$possible Lanczos potential'
$L^{(2)}_{abc} $ built around a spacelike Killing vector.  Although we believe
that it is straightforward   to show, in a generic manner, that the
defining equation (1.11a)  holds in vacuum --- by a proof formally analogous to
that given by  Novello and Velloso (1987) for the timelike case, but of course
without the physical interpretations --- we shall  instead avoid these long
calculations and use the result from the last section. 

We have  shown in  Appendix I,  that the Lanczos
differential condition  (1.4b) and the homogeneous wave equation (1.8) 
do hold for a generic potential $L^{(2)}_{abc} $ of this class.
Therefore  the result in the last section immediately verifies that 
{\it $L^{(2)}_{abc} $ is indeed a Lanczos potential in the Lanczos gauge ---
providing our background spacetime is vacuum, and does not fall into one of the
two exceptional cases.}

We cannot use this result for the Schwarzschild application (since the spacetime
is algebraically special), but we have already verified $L^{(2)}_{abc} $
directly for this case in part (iv) in Section 3.4.

We turn next to the other application of this class given in Dolan and Kim
(1994b).  Although the application, in Dolan and Kim (1994b), to the Bondi
space-time, makes no mention of any vacuum conditions being imposed on the
metric functions $U,V,g,b, $ we assume that it was intended for the vacuum
conditions to be understood. However, it will be interesting and instructive to
speculate whether this Lanczos potential is also valid for any {\it non-vacuum}
Bondi spacetime.\footnote{${}^{\dag} $}{The form of this new potential has been
proposed by analogy with the work of \NV (1987); although they considered the
more general perfect fluid spacetimes, this was not a necessary constraint on
their work and they stated that their results could be generalised further to
other non-vacuum spaces.  Since \DK construct their new potential around a
spacelike vector --- analogous to the timelike vector used by \NV --- the
results of \DK do not generalise {\it obviously in the same manner} to perfect
fluids; however, a generalisation to perfect fluids or other non-vacuum cases is
not explicitly ruled out.} We consider the metric as quoted in Dolan and Kim
(1994b), together with the vacuum equations in Di Prisco et al.  (1987).

First we consider the vacuum case. Since we know that the vacuum Bondi spacetime
has a very general form which does not fall into any of the two exceptional
cases in the last section, we can conclude from our result in the last section
that the potential proposed for the {\it vacuum} Bondi spacetime in Dolan and
Kim (1994b) is a genuine Lanczos potential in the Lanczos gauge for the Weyl
curvature tensor, up to a constant factor.

Turning now to the non-vacuum case we need to 
test directly whether each component of 
the Weyl curvature 
tensor $ \Psi_i $ calculated 
in the usual way from the NP equations, agrees with the corresponding component 
for the Weyl curvature tensor $\Psi^L_i $ calculated from  
the proposed
Lanczos potential. This is a long but straightforward
calculation by hand; but we did it very easily using Maple. 

We begin with the simplest component; using the explicit expressions for the
spin coefficients in Dolan and Kim (1994b) and the usual NP equations, (Newman
and Penrose (1962)), we obtain
$$\eqalign{\Psi_0  & = D\sigma - \sigma(\rho+\bar \rho)\cr
& = -e^{-4b}(rg_{11}-2rb_1g_1+2g_1)/r
}\eqno(5.1)$$
where the symbols have the meanings given in Di Prisco et al.  (1987).
On the other hand, using the  Weyl-Lanczos equations and the  explicit
expressions for the Lanczos potential (corrected by a factor of $(-1)$) 
 in  Dolan and Kim (1994b), we obtain
$$\eqalign{\Psi_0^L & = 2(D-\bar \rho)L_4 - 6\sigma L_1 \cr 
& = -e^{-4b}(rg_{11}-2rb_1g_1+2g_1+2b_1-rg_1^2)/r
}\eqno(5.2)$$
Clearly the $\Psi_0 $'s do not agree in general, but it is straightforward to
confirm that they agree when we make use of the field equation
$$\Phi_{00} = e^{-4b}(2b_1-rg_1^2)/r
\eqno(5.3)$$
in vacuum. (This is in fact the first
of the  main vacuum field
equations from   Di Prisco et al.  (1987)).

Carrying through the same calculations for all five  Weyl tensor components we
obtain 
$$\eqalign{\Psi_0 & =\Psi^L_0+\Phi_{00}\cr
\Psi_1 & =\Psi^L_1+\Phi_{01}\cr
\Psi_2 & =\Psi^L_2+{2\over 3}\Phi_{11}+{1\over 3}\Phi_{02}\cr
\Psi_3 & =\Psi^L_3+\Phi_{12}\cr
\Psi_4 & =\Psi^L_4+\Phi_{22}\cr}
\eqno(5.4)$$

Therefore we find that in order for all the Weyl-Lanczos equations to
be satisfied we need to use almost all the vacuum field equations; in 
particular we find that the most general form of the energy-momentum tensor is
$$ T_{ab}=\alpha (l_{(a}n_{b)} - x_a x_b) -\beta\Xi_{a}\Xi_b + \gamma g_{ab}
\eqno(5.5)$$ where $\Xi_a (= r e^{-g} \sin \theta(m_a-\bar m_a)/\sqrt 2 i)$ is
the spacelike Killing vector around which the Lanczos potential is constructed
and $x_a=(m_a+\bar m_a)/2$; $\{l_a,n_a, m_a,\bar m_a \}$ are the usual null
tetrad vectors given in Di Priso et al. (1987), and $\alpha ,\beta, \gamma $ are
arbitrary scalars.  Therefore we cannot use this form for the Lanczos potential
in many physically interesting non-vacuum Bondi spacetimes (e.g. perfect fluids,
radiation, electomagnetism). However, from the form in (5.5) we see we can
generalise to `vacuum' spacetimes where we permit a non-vanishing cosmological
constant, or equivalently to very special perfect fluids.

\

Finally we point out that the argument given above for the vacuum case left
the freedom of a constant factor; once we have  confirmed one of the 
components we can conclude that the value of the constant is unity. (Of course,
if we prefer not to use the general result in the previous section to establish
the vacuum case, we now have a direct proof for the special case of the
Bondi metric.)

\vfill\eject

\beginsection APPENDIX I

We consider first the class of  spacetimes and Lanczos potential considered 
in Lemma 3(ii) by  Novello and Velloso (1987),   in case (b) by  Dolan
and Kim (1994b), and summarised in (ii) in Section 3 of this paper. The
potential has the form
$$\eqalign{L_{abc}  ={1\over 3} (\sigma_{ca}v_b-\sigma_{ab}v_c) 
}$$
and using the properties given  in Novello and Velloso (1987), we can show by
direct substitution
$$\eqalign{L_{ab}{}^a & ={1\over 3} (\sigma^a{}_{a}v_b-\sigma_{ab}v^a) \cr & = 0
}\eqno(A.1a)$$
$$\eqalign{L_{ab}{}^c{}_{;c} & = {1\over 3}
(\sigma_{ca}{}^{;c} v_b-\sigma_{cb}{}^{;c}v_a+\sigma_{ca}
v_b{}^{;c}-\sigma_{cb}v_a{}^{;c}) \cr & = 
{1\over 3}
(\sigma_{ca}{}^{;c} v_b-\sigma_{cb}{}^{;c}v_a+\sigma_{ca}
(\sigma_b{}^{c}+\theta h_b{}^{c}/3) -\sigma_{cb}(\sigma_a{}^{c}+\theta
h_a{}^{c}/3) )
\cr & = 
{1\over 3}
(\sigma_{ca}{}^{;c} v_b-\sigma_{cb}{}^{;c}v_a)
\cr & = 
{2\over 9}
(\theta_{,a} v_b-\theta_{,b} v_a)}\eqno(A.1b)$$
Obviously the Lanczos algebraic gauge is always automatically satisfied, but
the Lanczos differential gauge is not always satisfied for all possible
spacetimes in this class.

 \

We consider next the new proposal in Dolan and Kim (1994b) --- the class of
vacuum spacetimes admitting a hypersurface-orthogonal space-like Killing vector
$\xi_a $. As pointed out in Dolan and Kim (1994b) the associated unit vector
$U_i= \xi_i/\xi$ satisfies
$$\eqalign{U^i{}_{;i}=0
}\eqno(A.2a)$$
   $$\eqalign{U_{(i;j)}(\delta ^i_a-U^iU_a) (\delta ^j_b-U^jU_b)=0
   }\eqno(A.2b)$$
   $$\eqalign{U_{i;j}=-A_iU_j \qquad\qquad\hbox{where}\qquad
A_i=U_{i;j}U^j=-(\ln\xi)_{,i} }\eqno(A.2c)$$
In addition, we note that
   $$\eqalign{U^iA_i=0
}\eqno(A.3a)$$
$$\eqalign{\square U_i=-U_iA_jA^j  
}\eqno(A.3b)$$
$$\eqalign{-\square \ln\xi=A^i{}_{;i}=U^i{}_{;ji}U^j+U^i{}_{;j}U^j{}_{;i}=0  
}\eqno(A.3c)$$
$$\eqalign{\square A_i=-(\square\ln\xi)_{,i}=0  
}\eqno(A.3d)$$
in vacuum.

\DK have proposed as a possible Lanczos potential
$$\eqalign{L^{(2)}_{ijk}= (A_iU_j-A_jU_i)U_k+{1\over 3}(A_ig_{jk}-A_j g_{ik})
}\eqno(A.4)$$
and this clearly satisfies the Lanczos algebraic gauge. 
By direct substitutions, using the properties listed above, we easily obtain
$$\eqalign{L^{(2)}_{ijk}{}^{;k}  =
0}\eqno(A.5)$$
and
$$\eqalign{\square L^{(2)}_{ijk}  = 0
}\eqno(A.6)$$
so the Lanczos differential gauge and the simple wave
equation are satisfied for the class of potentials given by $L^{(2)}_{ijk} $.

\vfill\eject 

\beginsection Acknowledgements.

This work was supported, in part, by the Swedish Natural Science Research
Council. 

\

\beginsection REFERENCES.

Andersson, F. and Edgar, S. Brian (1995). {\it The wave equations 
for the Weyl tensor/spinor and dimensionally dependent tensor identities.}
Preprint, Department of Mathematics,  University of Link\"oping, Sweden.

Ares de Parga G., et al (1989) {\it J.Math. 
Phys.,} {\bf 30,} 1294.

Atkins, W.K. and Davis, W.R. (1980). {\it Il Nuovo 
Cimento}, {\bf B59}, 116.

Bampi, F., and Caviglia, G. (1983). {\it Gen. Rel. 
Grav.},{\bf 15}, 375.

Bell, P and Szekeres P. (1972). {\it Int. J. 
Theor. Phys.,} {\bf  6,} 111.

Di Prisco, A., Herrera, L., Jimenez, J., Galina, 
V. and Ibanez, J. (1987). {\it J.Math.Phys.,} {\bf 
28,} 2692.

Dianyan, Xu (1986) {\it Phys. Rev. D}, {\bf 35,} 
769

Dolan, P. and Kim, C.W. (1994a). {\it Proc. R. 
Soc. Lond. A}, {\bf 447,} 557.

Dolan, P. and Kim, C.W. (1994b). {\it Proc. R. 
Soc. Lond. A}, {\bf 447,} 577.

Edgar, S. Brian (1994a).  {\it Mod. Phys. Lett. A, 
} {\bf 9},  479.

Edgar, S. Brian (1994b).  {\it G.R.G.}, {\bf 26}, 
329.

Edgar, S. Brian and H\"oglund A. (1995) {\it 
Dimensionally dependent identities for Curvature 
and  Lanczos tensors.} Preprint, Department of 
Mathematics, University of Link\"oping, Sweden.

Fulling, S A et al. (1992). {\it Class. Quant. 
Grav.}, {\bf 9,} 1151

Hammon K.S. and Norris L.K. (1993) {\it Gen. Rel. 
Grav.,} {\bf  25,} 55.

Harvey, A. (1995). {\it J. Math. Phys.}, {\bf 36,} 
356

H\"oglund, A. (1995) {\it "The Lanczos potential 
and its wave equation."}, Preprint, 
LiTH-MAT-Ex-95-08. Department of Mathematics, 
University of Link\"oping, Sweden.

Illge, R. (1988). {\it Gen. Rel. Grav.}, {\bf 20}, 
551.

Illge, R. (1995). {\it Private Communication.}

Jack, I. and Parker, L. (1987). {\it J. Math. 
Phys.}, {\bf 28,} 1137

Kramer, D., Stephani, H., Herlt, E., and MacCallum, M. A. H. (1980).
{\it Exact solutions of Einstein's field equations}, 
Cambridge University Press.

Lanczos, C. (1962). {\it  Rev. Mod. Phys.}, {\bf 
34}, 379.

L\'opez-Bonilla, J.L. et al  (1993) {\it Class. 
Quant. Grav.}, {\bf 10,} 2153

Lovelock, D. (1970). {\it Proc. Camb. Phil. Soc.}, 
{\bf 68,} 345

Maher, W.F. and  Zund, J.D. (1968). {\it Il Nuovo 
Cimento, Serie X,} {\bf  57 A,} 638.

Newman, E.T., and Penrose, R. (1962). {\it 
J.Math.Phys.,} {\bf 3,} 566.

Novello, M and Velloso, V. (1987). {\it Gen. Rel. 
Grav.}, {\bf 19}, 1251.

Parker L. and Christensen S.M. (1991). {\it 
MathTensor, } MathSolutions Inc., Chapel Hill.

Penrose, R.  and  Rindler, W. (1984). {\it Spinors 
and Spacetime Vols.1 and 2} (Cambridge University 
Press) .

Roberts, M.D. (1989). {\it Mod. Phys. Lett. A}, 
{\bf 4}, 2739. 

Taub, A.H. (1975). {\it Comp. \& Math. Appl.,} 
{\bf  1,} 377

Torres del Castillo, G.F. (1995) {\it 
J.Math.Phys.,} {\bf 36,} 195.

Zund, J.D. (1975). {\it Ann. Math.Pura Appl.,} 
{\bf  104,} 239.

\end